\documentclass[aps,twocolumn]{revtex4}

\usepackage{dcolumn}
\usepackage{bm}
\usepackage{graphicx}

\newcommand{\nc}{\newcommand}
\nc{\bc}{\begin{center}}
\nc{\ec}{\end{center}}

\nc{\noi}{\noindent}     
\nc{\eq}[1]{\mbox{Eq.~(\ref{#1})}}
\nc{\ba}{\begin{array}}
\nc{\ea}{\end{array}}
\nc{\bea}{\begin{eqnarray}}
\nc{\eea}{\end{eqnarray}}
\nc{\fig}[1]{\mbox{Fig.~\ref{#1}}}

\begin{document}

\title{Decay Processes in the Presence of Thin Superconducting Films}

\author{Per K. Rekdal $^{\, a,}$}
  \email{per.rekdal@uni-graz.at}
\author  {Bo-Sture K. Skagerstam$^{\, b,}$}
  \email{boskag@phys.ntnu.no}

\affiliation{ $^a$ Institut f\"ur Theoretische Physik, Karl-Franzens-Universit\"at Graz, Universit\"atsplatz 5, A-8010 Graz, Austria 
              \\
              $^b$ Complex Systems and Soft Materials Research Group, Department of Physics,
              The Norwegian University of Science and Technology, N-7491 Trondheim, Norway
            }

{\begin{abstract}
     \small 

     In a recent paper [Phys. Rev. Lett. {\bf 97}, 070401 (2006)] the  transition rate of magnetic spin-flip of a
     neutral two-level atom trapped in the vicinity of a thick superconducting body was studied.
     In the present paper we will  extend these considerations  
     to a situation with an atom at various distances from a dielectric film. 
     Rates for  the corresponding electric dipole-flip transition will also be considered.
     The rates for these atomic flip transitions can be
     reduced or enhanced, and in some situations they can even be completely suppressed.
     For a superconducting film or a thin film of a perfect conducting material various analytical
     expressions are derived that  reveals the dependence of the physical parameters at hand.
     %
     \\[5mm]
\noindent PACS numbers 34.50.Dy, 03.75.Be, 42.50.Ct  
\end{abstract}
}
\maketitle
%
%
%
\bc{
\section{INTRODUCTION}
\label{sec:introd}}
\ec
%
%
%
\vspace{-0.7cm}
     Harnessing the interactions of electromagnetic field and matter
     is one of the ultimate goals of atom optics. 
     One promising approach towards control of matter waves on small scales is to 
     trap and manipulate cold neutral atoms in microtraps near structures microfabricated on a surface,
     known as atom chips \cite{folman_02}.
     Magnetic traps on such atom chips are commonly generated either by microfabricated current-carrying
     wire \cite{folman_02} or by poled ferromagnetic films \cite{hinds_99,eriksson_04} attached to some
     dielectric or metallic body.
     However, the proximity of the atoms to the surface threatens to decohere the
     quantum state of the atoms through electromagnetic field fluctuations.
     This effect arises because the resistivity of the surface is always accompanied by field
     fluctuations as a consequence of the fluctuation-dissipation theorem.
     For an atom close to the surface of a dielectric body these fluctuating fields can be
     strong enough to drive magnetic and electric dipole transitions in the atom, as e.g.
     shown in recent experimental studies \cite{hinds_03,vuletic_04,harber_03}.
     If the atom is in a magnetic or electric trap, these flip transitions may lead to atom loss. 
     Such transitions are therefore most often undesirable, and we want to reduce them or even suppress them completely.

\begin{figure}[t]

\begin{picture}(0,0)(113,330)   

\includegraphics{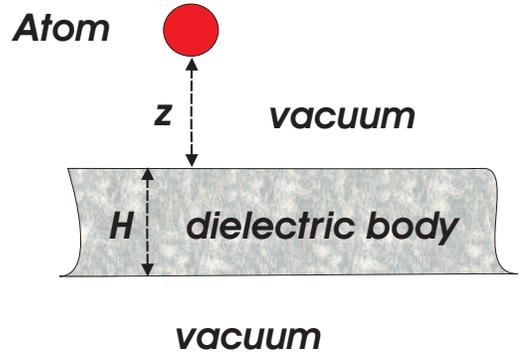}

\end{picture}

\vspace{5cm}

\caption{Schematic picture of the setup considered in our calculations. An atom inside a microtrap is 
         located in vacuum at a distance $z$ away from a dielectric slab with thickness $H$.
         Vacuum is on both sides of the slab. The dielectric slab can e.g. be a normal conducting metal or a
         superconducting metal.
         Upon making a magnetic spin-flip transition or an electric dipole-flip transition, the atom becomes more
         weakly trapped and is eventually lost.}
\label{geo_slab_fig}

\end{figure}

     In the present paper we intend to explicitly write down the flip rate for both of these types
     of transitions for the dielectric slab as shown in \fig{geo_slab_fig}. We will e.g. consider 
     a normal conducting slab as well as a superconducting slab, as described in  Ref.\cite{rekdal_06}.  
     To the best of our knowledge, this is not done for a general spin or dipole orientation, despite the fact
     that it in principle has been known for a long time (see e.g. Ref.\cite{knight_73}).

\vspace{1cm}

\bc{
\section{THEORY}
\label{sec:theory}
}\ec

\vspace{-1cm}

\bc{
\subsection{Magnetic Spin Transition}
}\ec
     We begin by considering an atom in an initial state $|i \rangle$ and trapped at position ${\bf r}_A= (0,0,z)$
     in vacuum, near a dielectric body. The rate $\Gamma_B $ of spontaneous and thermally stimulated magnetic spin-flip
     transition into a final state $|f\rangle$ has e.g. been derived in Ref. \cite{rekdal_04}, 
\bea \label{gamma_B_generel}
     \lefteqn{\Gamma_B = \, \mu_0 \,  \frac{2 \, (\mu_B g_S)^2}{\hbar} \; \sum_{j,k=1}^3 ~  
     S_j \, S_k^{\, *} } 
     \nonumber 
     \\ 
     && \; \times \;
     \mbox{Im}  \, [  \; \nabla \times \nabla \times
     \bm{G}({\bf r}_A  ,  {\bf r}_A  ,  \omega ) \; ]_{jk}  \;  ( \overline{n}_{\text{th}} + 1 ) \, ,
\eea

     \noi
     where we have introduced the dimensionless components $S_j \equiv   \langle f | \hat{S}_j/\hbar | i \rangle$ of the  electron spin operators $\hat{S}_j$      corresponding to the transition $|i\rangle \rightarrow |f\rangle$,
     with $j=x,y,z$.
     Here $\mu_B$ is the Bohr magneton, $g_S \approx 2$ is the electron spin $g$ factor,
     and $\bm{G}({\bf r}_A  ,  {\bf r}_A  ,  \omega )$ is the dyadic Green tensor of Maxwell's theory. 
     \eq{gamma_B_generel} follows from a consistent quantum-mechanical treatment of electromagnetic 
     radiation in the presence of absorbing bodies \cite{dung_00,henry_96}. 
     Thermal excitations of the electromagnetic field modes are accounted for by the factor 
     $( \overline{n}_{\text{th}} + 1 )$, where $\overline{n}_{\text{th}} = 1 / ( e^{\hbar \omega/k_{\text{B}} T}- 1 )$ is the Planck distribution
     giving the mean number of thermal photons per mode at frequency $\omega$ of the spin-flip transition. 
     Here $T$ is the temperature of the dielectric body, which is assumed to be in thermal equilibrium
     with its surroundings, and $k_B$ is Boltzmann's constant. The dyadic Green tensor is the unique
     solution to the Helmholtz equation 
\bea   \label{G_Helm}
   \nabla\times\nabla\times \bm{G}({\bf r},{\bf r}',\omega) - k^2
   \epsilon({\bf r},\omega) \bm{G}({\bf r},{\bf r}',\omega) = \delta( {\bf r} - {\bf r}' ) \bm{1} \, ,\nonumber\\
\eea

     \noi
     with appropriate boundary conditions. Here $k=\omega/c$ is the wavenumber in vacuum, $c$ is the speed of light 
     and $\bm{1}$ the unit dyad. The tensor $\bm{G}({\bf r},{\bf r}',\omega)$ contains all 
     relevant information about the geometry of the material and, through the electric permittivity
     $\epsilon({\bf r},\omega)$, about its dielectric properties.
     Due to causality, any complex dielectric function must in general obey the Kramers-Kronig relations. 
     Since we only consider non-zero frequencies in a suitable finite range,
     such dispersion relations will be of no concern in the present paper.

     The current density in a superconducting media is commonly described by the Mattis-Bardeen theory \cite{mattis_58}.
     Following  Ref. \cite{rekdal_06}, assuming non-zero frequencies $0 < \omega \ll \omega_g \equiv 2 \Delta(0)/\hbar$,
     where $\omega$ is the angular frequency and $\Delta(0)$ is the energy gap of the superconductor at zero temperature, 
     the current density is well described by means of a two-fluid model \cite{gorter_34,london_34_40}. 
     The dielectric function is in this case given by   \cite{rekdal_06}

\bea \label{eps_j}
  \epsilon(\omega) = 1 - \frac{1}{k^2 \lambda_L^2(T)} + i \, \frac{2}{k^2 \delta^2(T)} \, ,
\eea

    \noi
    where $\lambda_L(T) = e\sqrt{ m/\mu_0 \, n_s(T)}$ is the London penetration length
    and where $\delta(T) \equiv \sqrt{2/\omega \mu_0 \, \sigma_n(T)}$ is the skin depth associated with the
    normal conducting electrons. As usual, $\mu_0$ is the permeability of vacuum and $e$ is the elementary 
    charge. The total electron density $n_0(T)$ is constant and given by
    $n_0 = n_s(T) + n_n(T)$, where $n_s(T)$ and $n_n(T)$ are the electron densities
    in the superconducting and normal state, respectively, at a given temperature $T$.
    The optical conductivity corresponding to \eq{eps_j} is 
    $\sigma(T) = 2/\omega \mu_0 \delta^2(T) + i/\omega \mu_0 \lambda_L^2(T)$.    
    Above the transition temperature $T_c$, the dielectric function in \eq{eps_j}
    reduces to the well known Drude form. We also stress that the theory in this paper is particular
    to non-magnetic media.

    The rate $ \Gamma^{\, 0}_{ B } $ of a magnetic spin-flip transition for an atom in the unbounded
    free-space is well known (see e.g. Refs.\cite{dung_00}), with the result
\bea \label{gamma_0_B} 
     \Gamma^{\, 0}_{ B }  
   = {\bar  \Gamma}_{ B }  S^{\, 2} \; ,
\eea 
   \noi  with
\bea
     {\bar \Gamma}_{ B }  
   =  \, \mu_0  \, \frac{ ( \mu_B g_S )^2 }{3 \pi  \, \hbar} \, k^3 \; .
\eea 
    \noi
    Here we have introduced the dimensionless spin factor $S^{\, 2} \equiv S_x^{\, 2} + S_y^{\, 2} + S_z^{\, 2}$. 
    The unbounded free-space lifetime corresponding to this magnetic spin-flip rate is
    $\tau^{\, 0}_{ B } \equiv 1/\Gamma^{\, 0}_{ B }$.

    In the following we apply our model to the geometry shown in \fig{geo_slab_fig}, where an atom is located in vacuum
    at a distance $z$ away from a dielectric slab with thickness $H$. This slab is described by dielectric function as
    given by \eq{eps_j}. 
    The total magnetic transition rate 
\bea  \label{total_B}
  \Gamma_{ B } =  ( \Gamma^{\, 0}_{B} +  \Gamma^{\, \rm{slab}}_{B} ) \, ( \overline{n}_{\text{th}} + 1 ) \, ,
\eea
    \noi 
    can then be decomposed into a free part and a part purely due to the presence of the slab.
    The latter contribution for an arbitrary spin orientation is given by
\bea
    \Gamma^{\, \rm{slab}}_{B}  &=&  \label{gamma_B_gen_ja}
    2 \, {\bar \Gamma}^{\, 0}_{B}   \, 
    \bigg ( ~  (  S_x^{\, 2} \, + \, S_y^{\, 2}  ) \, I_{\|}   \, + \, S_z^{\, 2} \, I_{\perp} ~ \bigg )  ~ ,
\eea
    \noi
    with the atom-spin orientation dependent integrals 
\bea  \label{I_paral}
  I_{\|}    &=&
                \frac{3}{8}   
                {\rm Re} 
                \bigg ( 
        \int_{0}^{\infty} d q  \,  \frac{q}{\eta_0}  \, e^{ i \, 2 \eta_0 \, k z } \, 
        [  \,  {\cal C}_{N}(q)  -  \eta_0^2  \,  {\cal C}_{M}(q)   \, ]  \bigg ) \, , ~~~ \nonumber \\
\eea
   \noi and
\bea
   \label{I_perp}
   I_{\perp}  &=&   \frac{3}{4}  
                {\rm Re} 
                \bigg ( \,
                \int_{0}^{\infty}  d q  \, \frac{q^3}{\eta_0} \, e^{ i \, 2 \eta_0 \, k z } \, 
                C_{M}(q)  \, \bigg ) \, .
\eea
   \noi
   The scattering coefficients $C_{N}(q)$ and $C_{M}(q)$ are given by 
\bea  \label{C_N_33}
  {{\cal C}}_{N}(q)  =     r_p(q) ~ \frac{1 - e^{i \, 2 \, \eta(\omega) \, k H} }
                                         {1 \; - \; r_p^2(q)  \; e^{i \, 2 \, \eta(\omega) \, k H}} \, ,
\eea
  \noi  and
 \bea
   \label{C_M_33}
  {{\cal C}}_{M}(q)  =     r_s(q) ~ \frac{1 - e^{i \, 2 \, \eta(\omega) \, k H} }
                                         {1 \; - \; r_s^2(q)  \; e^{i \, 2 \, \eta(\omega) \, k H}} \, ,
\eea
   \noi
   with the electromagnetic field polarization dependent Fresnel coefficients
\bea \label{r_s_og_r_p}
  r_s(q)   = \frac{\eta_0 - \eta(\omega)}{\eta_0 + \eta(\omega)} ~~ , ~~
  r_p(q)   = 
                   \frac{\epsilon(\omega) \, \eta_0 - \eta(\omega)}
                        {\epsilon(\omega) \, \eta_0 + \eta(\omega)} \, .
\eea

    \noi 
    Here we have defined $\eta(\omega)  =  \sqrt{\epsilon(\omega) - q^2}$ and $\eta_0 = \sqrt{1 - q^2}$. 
    For a thick slab with $H=\infty$, the above equations are reduced to the results in Ref. \cite{henkel_99}.


\newpage

\bc{
\subsection{Electric Dipole Transition}
}\ec
%
%
%
%
%
      The previous section concerns magnetic field fluctuations. In this section we will consider
      electric field fluctuations.
      For an electrical dipole transition, the rate spontaneous and thermally stimulated decay 
      is given by (see e.g. Refs.\cite{dung_00}):
\bea \label{gamma_E_generel}
     \lefteqn{\Gamma_E = \, \mu_0 \,  \frac{2 \, \omega^2}{\hbar} \; \sum_{j,k=1}^3 ~  
     d_j \, d_k^{\, *} } 
     \nonumber 
     \\ 
     && \; \times \;
     \mbox{Im}  \, [  \;
     \bm{G}({\bf r}_A  ,  {\bf r}_A  ,  \omega ) \; ]_{jk}  \;  ( \overline{n}_{\text{th}} + 1 ) ~  ,
\eea
     \noi
     where $d_j \equiv \langle f|\hat{d}_j|i \rangle$, with $j=x,y,z$, is the matrix element of the atomic dipole operator
     $\hat{d}_j$ in the direction $j$,
     corresponding to the transition $|i \rangle \rightarrow |f \rangle$.

     Let us apply this model to the geometry shown in \fig{geo_slab_fig}, where an atom is located in vacuum at a
     distance $z$ away from a dielectric slab with thickness $H$. The total electric transition rate
\bea
    \Gamma_{E} = ( \Gamma^{\, 0}_{E} +  \Gamma^{\rm{slab}}_{E} ) \, ( \overline{n}_{\text{th}} + 1 ) \, ,
\eea
     \noi
     can then be decomposed into a free part and a part purely due to the presence of the slab.
     The latter contribution for an arbitrary dipole orientation is given by
\bea   \label{total_E}
    \Gamma^{\text{slab}}_{\, E} &=&   \
    2 \, {\bar \Gamma}^{\, 0}_{E} \, 
    \bigg ( ~  ( d_x^{\, 2} +  d_y^{\, 2} ) \, J_{\|}   \, + \,    d_z^{\, 2} \, J_{\perp} ~ \bigg )  \; ,
\eea
    \noi
    where we have introduced the dipole factor $d^{\, 2} \equiv d_x^{\, 2} +  d_y^{\, 2} +  d_z^{\, 2}$.
    The dipole orientation dependent integrals are
\bea  \label{J_paral}
  J_{\|}    &=& \nonumber
                \frac{3}{8}   
                {\rm Re} 
                \bigg ( 
        \int_{0}^{\infty} d q  \,  \frac{q}{\eta_0}  \, e^{ i \, 2 \eta_0 \, k z } \, 
        [  \,  {\cal C}_{M}(q)  -  \eta_0^2  \,  {\cal C}_{N}(q)   \, ]  \bigg ) \, ,
        \\ 
\eea
  \noi  and
\bea
   \label{J_perp}
   J_{\perp}  &=&                 \frac{3}{4}  
                {\rm Re} 
                \bigg ( \,
                \int_{0}^{\infty}  d q  \, \frac{q^3}{\eta_0} \, e^{ i \, 2 \eta_0 \, k z } \, 
                {\cal C}_{N}(q)  \, \bigg ) .
\eea
  \noi
  Here we have defined
\bea \label{gamma_0_E} 
\Gamma^{\, 0}_{E} =
  {\bar \Gamma }_{E}  
     \,  d^{\, 2} \; ,
\eea 
   \noi with
\bea  
   {\bar\Gamma }_{E}  
   =  \, \mu_0  \, \frac{ c^2 }{3 \pi  \, \hbar} \; k^3 \; ,
\eea 
   \noi
   i.e. the dipole-flip rate in unbounded vacuum for the electric dipole transition $|i \rangle \rightarrow |f \rangle$,
   with the corresponding free-space lifetime $\tau^{\, 0}_{E} \equiv 1/\Gamma^{\, 0}_{E}$. 
   We mention that \eq{gamma_0_E} is consistent with the definition in Refs. \cite{dung_00}.

\newpage

\bc{
\section{TWO LIMITING CASES}
\label{sec:perf_super}
}\ec

\subsection{Magnetic Spin Transition}

\subsubsection{ The Limit $\lambda_L(T) \ll \delta(T) , H, \lambda$ }

   Let us now consider a special case of the dielectric function in \eq{eps_j}. The superconducting term
   dominates over the normal conducting term provided that $\lambda_L(T) \ll \delta(T)$. If, in addition,
   $\lambda_L(T) \ll \lambda$, which holds true in practically all cases of interest,
   then we can neglect the unit term in \eq{eps_j}. 
   Here $\lambda = 2\pi/k$ is the wavelength associated to the magnetic spin-flip transition. 
   The dominant factor in the dielectric function is in this case real,
   and the main contribution to the integrals in Eqs. (\ref{I_paral}) and (\ref{I_perp}) occurs for values of 
   $q$ such that $0 \leq q \leq 1$. 
   For a slab with a thickness such that $\lambda_L(T) \ll H$, which also holds true in practically all cases
   of interest, the exponential functions in the scattering coefficients
   Eqs. (\ref{C_N_33}) and (\ref{C_M_33}) can be neglected. The scattering coefficients are then reduced to
   ${{\cal C}}_{N}(q) \approx r_p(q)$ and ${{\cal C}}_{M}(q) \approx r_s(q)$ for all relevant values of $q$.
   Furthermore, with the above mentioned assumptions, the Fresnel coefficients are reduced to $r_p(q) \approx 1$
   and $r_s(q) \approx - 1$.
   The integrals in Eqs. (\ref{I_paral}) and (\ref{I_perp}) can then be solved analytically. 
   The total magnetic spin-flip rate for an atom above a slab is then
\bea   \label{gamma_perf_B}
   \Gamma^{\, \text{pc}}_{B}  &\approx&   {\bar \Gamma}^{\, 0}_{B}   \, 
   ( \, \overline{n}_{\text{th}} + 1 \, ) 
   \\ \nonumber
   &\times& 
   \bigg [ \, S^{\, 2}  \, +  \,  \frac{3}{2} \, f_{\|}(kz)  \, ( S_x^{\, 2} + S_y^{\, 2} )
   \, + \, 3 \, f_{\perp}( k z ) ~ S_z^{\, 2}  ~~ \bigg ] \, ,
\eea

    \noi
    where we have defined

\bea
  f_{\|}( kz ) &\equiv&  \frac{\sin(  2 \, k z  )}{2 \, k z}  \, + \, f_{\perp}(kz) \, ,          
  \\
  f_{\perp}( kz ) &\equiv& \frac{ 2 \, k z \,  \cos(  2 \, k z  ) 
                   \, - \, 
                   \sin(  2 \, k z  )  }{(  2 \, k z )^3} \, .
\eea

    \noi
    Note that \eq{gamma_perf_B} is not valid for an arbitrary small thickness $H$ of the slab. 
    In the limit $\lambda_L(T) \rightarrow 0$, the magnetic spin-flip
    rate in \eq{gamma_perf_B} is exact.  This result  is consistent with Ref. \cite{henkel_99}.

    Let us now consider the near-field case $k z = \frac{2 \pi}{\lambda} \, z \ll 1$, which holds true in practically all cases of
    interest. The magnetic spin-flip rate is then reduced to

\bea \label{gamma_perf_c_and_n}
   \Gamma^{\text{pc} \, , \, \perp}_{B}  ~ \approx ~ \frac{ (2 \, k z)^2  }{10} ~  \Gamma^{\, 0}_{B} \, 
   ( \, \overline{n}_{\text{th}} + 1 \, )   \, ,
\eea

\begin{figure}[ht]

\begin{picture}(0,0)(165,455)   

\includegraphics{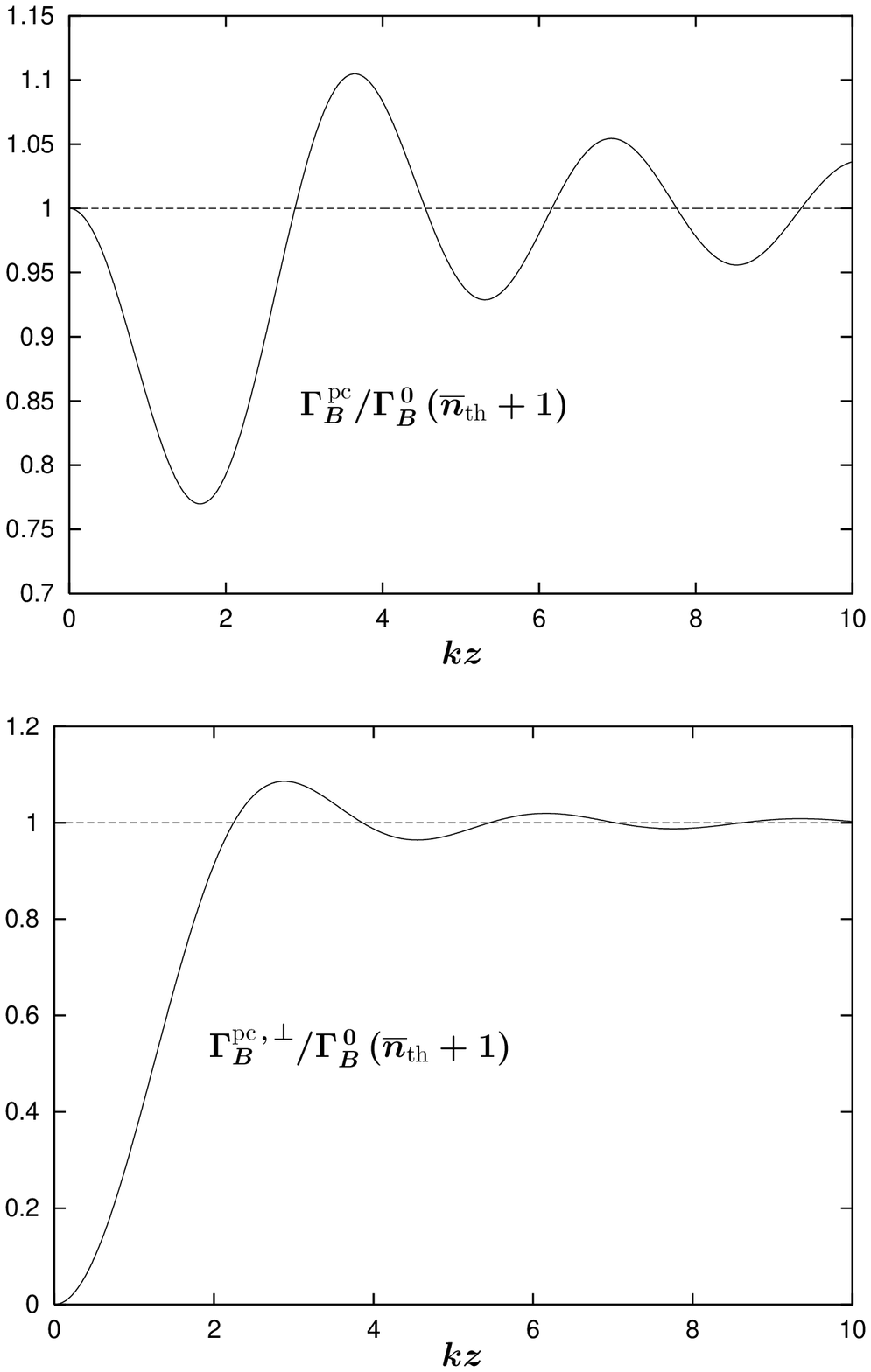}

\end{picture}

\vspace{14cm}

\caption{The lifetime $\Gamma^{\, \text{pc}}_{B}/\Gamma^{\, 0}_{B} \, ( \overline{n}_{\text{th}} + 1 )$
         according \eq{gamma_perf_B}. This fraction only depends on the product $kz$ and the atomic spin orientation. 
         {\it Upper figure}: The spin-orientation is the same as in Refs.\cite{scheel_05,rekdal_06}, i.e. 
         $|\langle f|\hat{S}_y|i\rangle|^2 = |\langle f|\hat{S}_z|i\rangle|^2$ and $\langle f|\hat{S}_x|i\rangle=0$.
         {\it Lower figure}:  The atomic spin is oriented perpendicular to the slab, i.e. 
         $\langle f|\hat{S}_x|i\rangle = \langle f|\hat{S}_y|i\rangle=0$. The spin-flip rate is completely
         suppressed for $kz=0$ in this case.
        }
\label{fig_kz}

\end{figure}

    \noi
    provided the atomic spin is oriented perpendicular to the slab, i.e. provided that
    $\langle f|\hat{S}_x|i\rangle = \langle f|\hat{S}_y|i\rangle = 0$. This result implies that, for an atom at
    the surface of the slab, i.e. $z=0$, there is no magnetic spin-flip at all (see lower graph in \fig{fig_kz}).
    The particular atomic spin orientation under consideration is the only one that can give zero spin-flip rate
    despite the presence magnetic field fluctuations. Furthermore, when the atomic spin is oriented parallel to the slab, 
    the magnetic spin-flip rate is

\bea \label{gamma_perf_B_paral}
   \Gamma^{\text{pc} \, , \, \|}_{B} 
                   ~ \approx ~ 2 ~  \Gamma^{\, 0}_{B} \, ( \, \overline{n}_{\text{th}} + 1 \, ) \, ,
\eea

    \noi
    for the near-field case $k z \ll 1$. This result shows that, in the near-field regime, the magnetic dipole-flip
    rate is twice the rate as compared to an atom in unbounded free-space, as e.g. pointed out in
    Ref.\cite{babiker_03}.
    In current atomic chip design (see e.g. Ref.\cite{hinds_03})
    the typical atomic frequency is $\omega/2 \pi = 560$ kHz and a typical atom-surface distance is $z=50 \, \mu$m.
    Hence, we have $k z \sim 10^{-7}$, i.e. far within the near-field condition $k z \ll 1$.

     In passing we also give the small $H$ expansion. For sufficiently small thickness of the slab, i.e.
     $H \ll \delta^2(T)/\lambda_L(T) , z , \lambda_L(T)$, the magnetic spin-flip rate is

\bea \nonumber
    \Gamma_{B} 
    &\approx&  \nonumber
    {\bar \Gamma}_{B} ~ ( \, \overline{n}_{\text{th}} + 1 \, )
    \\ \nonumber
    &\times&
    \bigg [ \, S^{\, 2} \, + \, ( S_x^{\, 2} + S_y^{\, 2} ) \, 
                         \frac{3}{64} ~   \frac{1}{( \, k \, \delta(T) \, )^2} \, \frac{1}{k \, \lambda_L(T)} \, ( \, \frac{H}{z} \, )^2  ~
  \\
  &&  
  ~~~~~~\, + ~ S_z^{\, 2} ~ \frac{3}{64} ~ \frac{k \, \lambda_L(T)}{( \, k \, \delta(T) \, )^2} \, 
  \frac{k H}{( \, k z \, )^3}  ~ \bigg ] \, . 
\eea

    \noi
    This  expression for the magnetic spin-flip rate as it is only valid for thicknesses of the slab
    smaller than the London penetration length.

\subsubsection{ The Limit $\delta(T) \ll \lambda_L(T) , H, \lambda, z$ }

    Let us now consider the case when the dielectric function as given by \eq{eps_j}
    is dominated by the normal conducting term, i.e. when $\delta(T) \ll \lambda_L(T)$ and 
    $\delta(T) \ll \lambda$.
    The dominant factor in the dielectric function is in this case imaginary, corresponding to the well known Drude form,
    and the main contribution to the integrals in Eqs. (\ref{I_paral}) and (\ref{I_perp}) is for values of $q$ such that
    $q \lesssim 1/kz$. 
    If, in addition, $\delta(T) \ll z$ then the exponential functions in the scattering coefficients in
    Eqs. (\ref{C_N_33}) and (\ref{C_M_33}) are negligible for all values $q \lesssim 1/kz$ provided that
    $\delta(T) \ll H$. 
    The scattering coefficients are then reduced to ${{\cal C}}_{N}(q) \approx 1$ and ${{\cal C}}_{M}(q) \approx -1$,
    i.e. the same result as in the last subsection.
    Hence, the conditions $\delta(T) \ll \lambda_L(T) , H, \lambda, z$ and $\lambda_L(T) \ll \delta(T) , H, \lambda$
    give the same results.
    %
    %
    In particular, for the perfect normal conducting limit, i.e. $\delta(T) \rightarrow 0$, the magnetic spin-flip
    rate as given by \eq{gamma_perf_B} is exact.  
    Note that the perfect normal conducting limit is only valid for the case 
    $\delta(T) \ll \lambda_L(T) , H, \lambda,z$, which e.g. means that the slab can not be arbitrarily thin.
    It also means that, in contrast to the case as described in last subsection, the atom-surface distance $z$ 
    can not be chosen arbitrary small.
    %

    We close this section by mentioning that, following the two-fluid model \cite{gorter_34,london_34_40} and
    using the Gorter-Casimir temperature dependence \cite{gorter_34}, the limit $\delta(T) \rightarrow 0$
    is obtained for $T \rightarrow 0$, as e.g. described in Ref. \cite{rekdal_06}.

\subsection{Electric Dipole Transition}

   \noi
   The only difference between the integrals in Eqs. (\ref{J_paral}-\ref{J_perp}) and 
   Eqs. (\ref{I_paral}-\ref{I_perp}) is the position of the scattering coefficients ${{\cal C}}_{N}(q)$
   and ${{\cal C}}_{M}(q)$. Hence, the correction to the vacuum dipole-flip rate corresponding to electric field
   fluctuations for the two limits as mentioned above is in general opposite in sign as compared to that of the
   magnetic spin-flip case. This was also pointed out in Ref.\cite{knight_73}. It can be understood in physical terms,
   since the electric and magnetic field are perpendicular. Hence, if $\delta(T) \ll \lambda_L(T) , H, \lambda, z$ or
   $\lambda_L(T) \ll \delta(T) , H, \lambda$ then the total electric dipole-flip rate
   for an atom above a slab is given by

\bea    \nonumber
   \Gamma^{\, \text{pc}}_{E}  &\approx&  {\bar \Gamma}_{E}   \, 
   ( \, \overline{n}_{\text{th}} + 1 \, ) 
   \\ 
   &\times&  \label{gamma_perf_E}
   \bigg [ \, d^{\, 2}  \, -  \,  \frac{3}{2} \, f_{\|}(kz)  \, ( d_x^{\, 2} + d_y^{\, 2} ) 
   \, - \, 3 \, f_{\perp}(kz) \, d_z^{\, 2} \; \bigg ] \, .\nonumber \\
\eea

      \noi
      This equation is consistent with the results in Ref. \cite{babiker_03}.

      Let us again consider the near-field case $k z \ll 1$. The electric dipole-flip rate is then reduced to

\bea \label{gamma_perf_E_paral}
         \Gamma^{\text{pc} \, , \, \|}_{E} 
         \approx \frac{ (2 \, k z)^2  }{5} ~ \Gamma^{\, 0}_{E} \, ( \, \overline{n}_{\text{th}} + 1 \, ) \, ,
\eea

    \noi
    provided that the atomic dipole is oriented parallel to the slab, i.e. provided that
    $\langle f|\hat{d}_z|i\rangle = 0$. This result implies that, for an atom at the
    surface of the slab, i.e. $z=0$, there is no electric dipole-flip at all. The particular atomic
    dipole orientation under consideration is the only one that can give zero dipole-flip rate
    despite the presence of electric field fluctuations. Furthermore, when the atomic spin is oriented
    perpendicular to the slab, the electric dipole-flip rate is

\bea \label{gamma_perf_E_perp}
   \Gamma^{\text{pc} \, , \, \perp}_{E} 
                ~ \approx ~ 2 ~  \Gamma^{\, 0}_{E} \, ( \, \overline{n}_{\text{th}} + 1 \, ) \, ,
\eea

   \noi
    for the near-field case $k z \ll 1$. This result was pointed out by Babiker in Ref. \cite{babiker_03}.

    To summarize, in the present paper we have reported results on the magnetic as well as electric
    decay properties of a neutral two-level atom trapped in the vicinity of a dielectric body.
    For a slab with vacuum on both sides (see \fig{geo_slab_fig}),
    we have obtained  the flip rate for both of these types of transitions, for
    any spin or dipole orientation. 
       The expression for the electric and magnetic transition rate can be solved exactly in two limiting cases, i.e. in
    the small skin depth limit for normal conducting metals and in the small
    London length limit for superconductors. In these limits, the correction
    to the vacuum rate for an electric dipole transition is opposite in sign as compared to that of a
    magnetic spin transition. These results are consistent with well known results, e.g. Refs. \cite{henkel_99}.

\begin{center}
{ \bf ACKNOWLEDGMENT }
\end{center}
%

     This work has been supported in part the Norwegian University of Science and Technology (NTNU) and 
     by the Austrian Science Fund (FWF). One of the authors (P.K.R.) wishes to thank Prof. Ulrich Hohenester at the 
     Karl-Franzens-Universit\"at Graz for warm hospitality while the present work was completed.


\end{document}